\documentclass[onecolumn,floatfix,prd,aps,10pt]{revtex4-2}
\usepackage[applemac]{inputenc}
\usepackage[T1]{fontenc}
\pdfoutput=1
\usepackage{enumerate}
\usepackage{braket}
\usepackage{amsmath}
\usepackage{subfigure}
\usepackage{lipsum}
\usepackage{amsfonts}
\usepackage{amssymb, mathrsfs}
\usepackage{bm}
\usepackage[pdftex]{graphicx}
\usepackage{footnote}
\usepackage{hyperref}  
\hypersetup{colorlinks=true,allcolors=blue}
\usepackage{hypcap}   
\usepackage{bookmark}
\usepackage{dsfont,mathrsfs,color,url,verbatim,booktabs}
\usepackage{booktabs}
\usepackage{tabularx}
\usepackage{threeparttable}

\def\beq{\begin{equation}}
	\def\eeq{\end{equation}}
\def\bsp{\begin{split}}
	\def\esp{\end{split}}
\def\bea{\begin{eqnarray}}
	\def\eea{\end{eqnarray}}
\def\ba{\begin{array}}
	\def\ea{\end{array}}

\def\l.{\left.}
\def\r.{\right.}

\def\part{\partial}

\begin{document}

\title{Chaplygin and Polytropic Kantowski--Sachs Solutions in Teleparallel $F(T)$ Gravity}
\author{Alexandre Landry\\
Department of Mathematics and Statistics, Dalhousie University, Halifax, NS B3H 3J5, Canada\\
\texttt{a.landry@dal.ca}}

\begin{abstract}
A covariant reconstruction framework for Kantowski--Sachs (KS) geometries sourced by Chaplygin-type and polytropic fluids in teleparallel $F(T)$ gravity is developed using the coframe--spin-connection formalism and the teleparallel invariant approach.  The matter sector is modeled by nonlinear equations of state, including the generalized Chaplygin gas $p=-A/\rho^{\alpha}$ and a polytropic law $p=K\rho^{\Gamma}$.  The corresponding conservation laws determine the dependence of the fluid density on the anisotropic KS volume $V=A_2A_3^2$.  These source scalings are then inserted into the symmetric part of the covariant teleparallel field equations and used to reconstruct the functional form of $F(T)$ directly from the KS dynamics.  Power-law and exponential ans\"atze generate distinct invariant reconstruction branches.  In the power-law sector, the Chaplygin fluid produces mixed constant-plus-power source terms, while the polytropic sector generates density powers controlled by the polytropic index.  In the exponential sector, the natural reconstruction variable is the shifted invariant $X=T_0-T$, leading to shifted teleparallel de Sitter branches.  The reconstructed models are interpreted as local anisotropic cosmological sectors and, for contracting angular KS scale factors, as local Kantowski--Sachs black-hole-interior reconstruction branches. The analysis is local and branch-dependent; leading-order viability is assessed through \(F_T>0\) and \(F_{TT}>0\), while a complete perturbative stability analysis is left for future work. The reconstruction is entirely driven by nonlinear matter conservation laws, thereby reversing the standard reconstruction strategy in which the gravitational Lagrangian is prescribed a priori.

\end{abstract}

\maketitle

\section{Introduction}

Teleparallel gravity provides a geometric formulation of gravitation in which the gravitational interaction is encoded in torsion rather than curvature \cite{AldrovandiPereira2013,KrssakSaridakis2016,Cai2016,Bahamonde2023}.  In the covariant formulation, the fundamental variables are a coframe $h^a{}_{\mu}$ and an inertial spin connection $\omega^a{}_{b\mu}$, which together define a curvature-free but torsionful geometry \cite{KrssakPereira2015,Golovnev2017,Krssak2019}.  This covariant formulation provides the geometric foundation for
modern teleparallel gravity, allowing modified torsional theories to
be formulated without sacrificing local Lorentz invariance while
maintaining a clear separation between inertial and gravitational
degrees of freedom \cite{KrssakSaridakis2016,Krssak2019,Golovnev2017}.

Modified teleparallel $F(T)$ gravity is a nonlinear extension of the teleparallel equivalent of general relativity (TEGR), obtained by replacing the torsion scalar $T$ by a function $F(T)$ in the gravitational action \cite{FerraroFiorini2007,Cai2016,Bahamonde2023}.  Because $F(T)$ models modify the torsional sector directly, their dynamics can differ significantly from curvature-based modified gravity, especially in anisotropic and inhomogeneous settings \cite{FerraroFiorini2007,Bahamonde2023,Capozziello2011}. Within this framework, invariant classification methods based on the coframe--spin-connection pair provide a natural geometric language for organizing physically inequivalent teleparallel solutions and their associated reconstruction branches \cite{Krssak2019,Coley2020,ColeyLandry2024,LandryKS2024}.  Kantowski--Sachs (KS) geometries provide a particularly useful testing ground because they are homogeneous but anisotropic, possess a natural spherical two-sector, and arise both in anisotropic cosmology and in black-hole-interior descriptions \cite{KantowskiSachs1966,RyanShepley1975,EllisMacCallum1969,HawkingEllis1973,Stephani2003}.

Although teleparallel gravity has been extensively investigated in
homogeneous and isotropic Friedmann--Lema\^itre--Robertson--Walker
(FLRW) cosmologies, anisotropic spacetimes remain considerably less explored than their FLRW counterparts, particularly within covariant $F(T)$ gravity
\cite{LeonRoque2014,Cai2016,Bahamonde2023,Rodrigues2015,Dimakis2023,Millano2024}.
Among these geometries, the Kantowski--Sachs (KS) spacetime occupies a
special position because it simultaneously describes homogeneous
anisotropic cosmologies and the local interior geometry of
spherically symmetric black holes
\cite{KantowskiSachs1966,RyanShepley1975,HawkingEllis1973,Stephani2003}.
Consequently, KS models constitute an ideal laboratory for
investigating how nonlinear torsional modifications interact with
anisotropic geometries while remaining analytically tractable.
The covariant coframe--spin-connection (CSC) formalism, together with
the teleparallel invariant classification, is particularly well
suited to this purpose because it preserves local Lorentz covariance
and provides a systematic classification of inequivalent teleparallel
branches
\cite{KrssakPereira2015,Golovnev2017,Krssak2019,Coley2020,ColeyLandry2024,LandryKS2024,HoogenForance2024}.

Nonlinear cosmological fluids have attracted considerable attention as
effective descriptions of dark energy, dark matter, and unified
dark-sector scenarios \cite{Copeland2006,Bamba2012,Nojiri2011,Harko2011,Frieman2008}.
Among the most widely studied phenomenological models is the generalized
Chaplygin gas, which naturally interpolates between a dust-dominated
regime at early times and a cosmological-constant-like phase at late
times
\cite{Kamenshchik2001,Bento2002,Bilic2002,Gorini2003,Bouhmadi2008}.
Modified Chaplygin models further extend this behaviour through an
additional barotropic contribution and have been widely investigated
in cosmology and modified gravity \cite{Debnath2004,Bertolami2004,Jamil2011}.
Similarly, polytropic equations of state constitute one of the standard
effective descriptions of relativistic stellar structure,
compact objects,
self-gravitating fluids,
and several cosmological scenarios,
thereby providing physically realistic nonlinear matter sources with direct
astrophysical relevance \cite{Tooper1964,Tooper1965,Chavanis2008,Karami2011,Glendenning2000,Haensel2007,Horedt2004,Lattimer2016}. These nonlinear fluids therefore constitute physically motivated sources for investigating covariant teleparallel gravity beyond the standard linear perfect-fluid approximation. Recent observational analyses have shown that nonlinear
Chaplygin-type models remain useful phenomenological
parameterizations for testing unified dark-sector scenarios,
although their parameter space is increasingly constrained by
modern cosmological observations
\cite{Bertolami2004,Planck2018}. Beyond cosmology, Chaplygin-type equations of state have also been
employed in relativistic stellar models, compact objects, wormhole
geometries, and anisotropic self-gravitating configurations, making
them versatile effective descriptions across a wide range of
gravitational systems \cite{Lobo2006,Rahaman2012}.

Most previous investigations involving Chaplygin or polytropic matter
in modified gravity prescribe the gravitational action
$a~priori$ and subsequently solve the corresponding field equations.
By contrast, the reconstruction strategy adopted here follows the
inverse reconstruction approach \cite{LandryEMKS2026}. This philosophy parallels inverse reconstruction methods
developed in other modified-gravity frameworks,
but here the invariant CSC formulation allows the
reconstruction to remain manifestly covariant
throughout the procedure
\cite{Capozziello2011,Cai2016,Bahamonde2023,Nojiri2011,Bamba2012}. The nonlinear conservation law first fixes the evolution of the matter
density through the KS volume factor, after which the reduced
teleparallel field equations are used to infer the gravitational
Lagrangian \(F(T)\). In this way, the matter equation of state fixes
the admissible invariant source sector before the gravitational
Lagrangian is reconstructed \cite{Cai2016,Bahamonde2023,ColeyLandry2024}.

The present work extends the covariant electromagnetic KS reconstruction program to nonlinear perfect-fluid sources \cite{LandryEMKS2026}.  In the electromagnetic case, Maxwell conservation laws impose restricted source scalings and thereby act as dynamical selection rules for admissible reconstruction branches \cite{LandryEMKS2026}.  Here, the role of the source constraint is instead played by nonlinear fluid conservation laws associated with Chaplygin-type and polytropic equations of state.  Chaplygin gases and their generalized forms are widely used as effective models for dark-energy--dark-matter unification and late-time accelerated cosmology \cite{Kamenshchik2001,Bento2002,Bilic2002,Gorini2003}.  Polytropic fluids are also standard effective matter models in relativistic astrophysics and cosmology, especially in compact-object modeling and phenomenological cosmological reconstruction \cite{Tooper1964,Chavanis2008,Karami2011}.

From this viewpoint, Chaplygin and polytropic fluids do not merely
provide alternative matter sources. Rather, their nonlinear equations
of state modify the source sector of the reduced teleparallel field
equations and therefore generate reconstruction branches that are
structurally distinct from electromagnetic, scalar-field, or standard
barotropic-fluid sectors \cite{LandryEMKS2026,LandryKS2024,Cai2016,Bahamonde2023}.

The aim of the present work is to construct a local invariant reconstruction scheme for KS geometries sourced by Chaplygin and polytropic fluids in covariant $F(T)$ gravity \cite{Coley2020,ColeyLandry2024}.  The reconstruction strategy is not to impose a predefined $F(T)$ action, but rather to determine the functional form of $F(T)$ from the reduced teleparallel field equations after the matter conservation law has fixed $\rho(V)$, where
\begin{equation}
V(t)=A_2(t)A_3^2(t)
\end{equation}
is the KS volume scale factor. The principal contributions of the present work are:
\begin{itemize}
\item a covariant CSC treatment of KS geometries sourced by nonlinear Chaplygin and polytropic fluids;
\item a reconstruction algorithm using nonlinear conservation laws rather than Maxwell flux constraints;

\item a derivation of explicit power-law and exponential reconstruction
branches in terms of $T$ or $X=T_0-T$;

\item a local interpretation of the resulting sectors as anisotropic cosmological or KS-interior-like branches;

\item a leading-order viability assessment based on $F_T>0$ and $F_{TT}>0$;

\item explicit treatment of generalized and modified Chaplygin gases;

\item a unified reconstruction framework encompassing generalized Chaplygin, modified Chaplygin, polytropic, and barotropic fluid sectors.

\end{itemize}

From a broader perspective, the present work extends the recent
covariant electromagnetic KS reconstruction program by replacing
Maxwell conservation laws with nonlinear fluid conservation equations \cite{LandryEMKS2026}. This generates genuinely new invariant reconstruction sectors whose structure is directly controlled by the matter equation of state.

The reconstruction developed throughout this work is local and
branch-dependent. Accordingly, the reconstructed functions should be
interpreted as local realizations of admissible invariant teleparallel
branches associated with fixed CSC sectors, rather than as unique
fully consistent global cosmological or compact-object solutions. 
In particular, the interpretation of contracting KS geometries as
black-hole-interior-like sectors remains local and requires matching
to an exterior spacetime, horizon analysis, and a complete
perturbative stability study before physically realistic compact
objects can be obtained.
Similarly, the viability conditions employed throughout the paper,
including $F_T>0$ and $F_{TT}>0$, should be regarded as necessary
leading-order criteria.
A comprehensive assessment of the physical viability of the reconstructed
models will require cosmological perturbations, dynamical-system
analyses, and observational constraints, which are left for future
work \cite{Bahamonde2018,Planck2018}. The framework developed here may also serve as a basis for future
extensions involving scalar fields, non-minimal couplings,
teleparallel New General Relativity (NGR), $F(T,B)$ gravity, and other generalized teleparallel theories, thus extending the invariant reconstruction program developed for covariant teleparallel gravity \cite{Hayashi1979,Beltran2023,Bahamonde2015}. The present analysis therefore complements recent invariant
classification programs by extending them from vacuum and
electromagnetic sectors to nonlinear matter sources. The remainder of the paper is organized as follows.
Section \ref{sect2} summarizes the covariant $F(T)$ field equations
and the Kantowski--Sachs geometry.
Section \ref{sect3} derives the nonlinear conservation laws.
Sections \ref{sect4}--\ref{sect5} develop the power-law and exponential
reconstruction branches.
Section \ref{sect6} discusses local de Sitter and interior-like sectors,
Section \ref{sect7} presents necessary leading-order viability conditions,
and Section \ref{sect8} summarizes the main conclusions.

\section{Teleparallel Field Equations and Kantowski--Sachs Geometry}\label{sect2}

\subsection{Covariant Teleparallel $F(T)$ Gravity}

We use Greek indices for spacetime components and Latin indices for
orthonormal tangent-space components. The tangent-space metric is
\(\eta_{ab}=\mathrm{diag}(1,-1,-1,-1)\), and the spacetime metric is
constructed from the coframe according to
\begin{equation}
g_{\mu\nu}=\eta_{ab}h^a{}_\mu h^b{}_\nu .
\end{equation}
The inertial spin connection \(\omega^a{}_{b\mu}\) is assumed to be
flat, so that \(R^a{}_{b\mu\nu}(\omega)=0\) \cite{AldrovandiPereira2013,KrssakSaridakis2016,Krssak2019}.

The action for minimally coupled matter is
\begin{equation}
S=\int d^4x\,h\left[\frac{1}{2\kappa}F(T)+\mathcal{L}_{m}\right],
\label{action}
\end{equation}
where $h=\det(h^a{}_{\mu})$, $T$ is the torsion scalar, $F_T=dF/dT$, $F_{TT}=d^2F/dT^2$, and $\kappa=8\pi G$. The torsion tensor is defined by
\begin{equation}
T^a{}_{\mu\nu}
=
\partial_\mu h^a{}_\nu-\partial_\nu h^a{}_\mu
+\omega^a{}_{b\mu}h^b{}_\nu
-\omega^a{}_{b\nu}h^b{}_\mu ,
\end{equation}
and the torsion scalar is
\begin{equation}
T=S_a{}^{\mu\nu}T^a{}_{\mu\nu},
\end{equation}
where \(S_a{}^{\mu\nu}\) denotes the teleparallel superpotential \cite{AldrovandiPereira2013,Cai2016,Bahamonde2023}. Throughout this work we adopt the sign conventions of Refs.~\cite{AldrovandiPereira2013,KrssakSaridakis2016,Krssak2019}, ensuring consistency with the covariant teleparallel literature. Here \(S_a{}^{\mu\nu}\) is constructed from the contortion and torsion vector in the standard teleparallel way.

For completeness we first recall the covariant field equations in
their compact tensorial form before specializing them to the
Kantowski--Sachs geometry. Varying the action with respect to the coframe gives the covariant $F(T)$ field equations \cite{KrssakSaridakis2016,Krssak2019}:
\begin{equation}
\kappa \Theta^{\mu}{}_{a}
=h^{-1}F_T\partial_{\nu}\left(hS_a{}^{\mu\nu}\right)
+F_{TT}S_a{}^{\mu\nu}\partial_{\nu}T
+\frac{F}{2}h_a{}^{\mu}
-F_TT^b{}_{a\nu}S_b{}^{\nu\mu}
-F_T\omega^b{}_{a\nu}S_b{}^{\nu\mu} .
\label{generalFE}
\end{equation}
The symmetric and antisymmetric parts can be written schematically as
\begin{align}
\kappa\Theta_{(ab)}&=F_T\,\mathring{G}_{ab}+F_{TT}S^{\mu}{}_{(ab)}\partial_{\mu}T
+\frac{\eta_{ab}}{2}\left(F-TF_T\right),
\label{SFE}\\
0&=F_{TT}S^{\mu}{}_{[ab]}\partial_{\mu}T .
\label{AFE}
\end{align}
The antisymmetric equations are purely teleparallel consistency
conditions: they do not introduce additional matter equations, but
restrict the admissible coframe--spin-connection branches in nonlinear
\(F(T)\) gravity and disappear identically in the
TEGR limit \(F(T)=T+\mathrm{const.}\), where \(F_{TT}=0\) \cite{KrssakSaridakis2016,Golovnev2017,Krssak2019}.

For minimally coupled metric matter, the matter
energy--momentum tensor satisfies the standard Levi--Civita conservation
law,
\begin{equation}
\mathring{\nabla}_\mu \Theta^{\mu\nu}=0 .
\end{equation}
This conservation law follows from diffeomorphism invariance of the
matter action and is independent of the explicit functional form of
\(F(T)\) \cite{Krssak2019,Bahamonde2023}.

\subsection{Kantowski--Sachs Coframe and Spin Connection}

We use the orthonormal KS coframe
\begin{equation}
h^a{}_{\mu}=\mathrm{Diag}\left[1,A_2(t),A_3(t),A_3(t)\sin\theta\right],
\label{KScoframe}
\end{equation}
which gives
\begin{equation}
ds^2=dt^2-A_2^2(t)dr^2-A_3^2(t)d\Omega^2.
\end{equation}
The functions \(A_2(t)\) and \(A_3(t)\) describe respectively the radial
and angular scale factors of the anisotropic KS geometry. The choice of
the cosmic-time gauge \(h^0{}_t=1\) is standard in homogeneous
cosmological models and leaves two independent anisotropic expansion
rates \cite{KantowskiSachs1966,RyanShepley1975,EllisMacCallum1969}.

The compatible inertial spin connection is chosen as
\begin{equation}
\omega^2{}_{34}=-\omega^2{}_{43}=\delta,\qquad
\omega^3{}_{44}=-\frac{\cos\theta}{A_3\sin\theta},
\label{spinconnection}
\end{equation}
with $\delta=\pm1$. The spin connection removes inertial contributions associated with the
spherical sector and ensures that the torsion scalar and field equations
are computed in a covariant teleparallel frame. The discrete parameter
\(\delta=\pm1\) labels the two orientation choices of the angular
teleparallel frame. This is the KS CSC branch used in the invariant teleparallel classification program \cite{KrssakPereira2015,Krssak2019,Coley2020,ColeyLandry2024,LandryKS2024,HoogenForance2024}.

For later use, we introduce the anisotropic Hubble rates
\begin{equation}
H_2=\frac{\dot A_2}{A_2},\qquad
H_3=\frac{\dot A_3}{A_3}.
\end{equation}
Then the torsion scalar becomes
\begin{equation}
T=2H_3(H_3+2H_2)-\frac{2}{A_3^2}.
\label{torsionKS}
\end{equation}
The torsion scalar depends on both anisotropic expansion rates and on
the intrinsic curvature of the spherical two-surfaces, thereby encoding
the anisotropic dynamics of the KS teleparallel sector.

\subsection{Reduced Field Equations and Isotropic Fluid Source}

The matter source is written in the orthonormal frame as
\begin{equation}
\Theta^a{}_b=\mathrm{diag}(\rho,-p_r,-p_t,-p_t),
\end{equation}
where \(p_r\) and \(p_t\) denote the radial and angular pressures.

The independent symmetric field equations are
\begin{align}
\kappa\rho&=-\frac{1}{2}\left(F-TF_T\right)
+F_T\left[\frac{1}{A_3^2}+2\frac{\dot A_2}{A_2}\frac{\dot A_3}{A_3}+\frac{\dot A_3^2}{A_3^2}\right],
\label{rhoeq}\\
-\kappa p_r&=\frac{1}{2}\left(F-TF_T\right)
-2 F_{TT}\dot{T} \frac{\dot A_3}{A_3}
-F_T\left[\frac{1}{A_3^2}+2\frac{\ddot A_3}{A_3}+\frac{\dot A_3^2}{A_3^2}\right],
\label{preq}\\
-\kappa p_t&=\frac{1}{2}\left(F-TF_T\right)
-F_{TT}\dot{T} \left(\frac{\dot A_3}{A_3}+\frac{\dot A_2}{A_2}\right)
-F_T\left[\frac{\ddot A_2}{A_2}+\frac{\ddot A_3}{A_3}+\frac{\dot A_2}{A_2}\frac{\dot A_3}{A_3}\right].
\label{pteq}
\end{align}
These equations follow directly from the covariant
coframe variation after specialization to the KS CSC
branch and coincide with the equations derived in
Ref.~\cite{LandryEMKS2026} in the appropriate limit. For the nonlinear perfect-fluid sectors considered below, we take
\begin{equation}
p_r=p_t=p(\rho),
\end{equation}
so that the matter source is isotropic in the orthonormal frame. Although the Chaplygin and polytropic sources considered below are
isotropic in the orthonormal frame, the KS geometry remains anisotropic
because the independent geometric sectors of the field equations impose
different constraints on \(A_2(t)\), \(A_3(t)\), and \(F(T)\).

Combining the first two equations gives a reduced invariant relation of the form
\begin{equation}
\partial_t\ln F_T=\frac{\kappa(\rho+p_r)+F_T\,\mathcal{G}_{rp}}{2(\dot A_3/A_3)},
\label{generalreduced}
\end{equation}
where \(G_{rp}\) denotes the purely geometric difference between the
radial-pressure and energy-density combinations appearing in
Eqs.~\eqref{rhoeq}--\eqref{preq}, after isolating the \(\dot F_T\) contribution. Equation~\eqref{generalreduced} is the reduced reconstruction relation used throughout the rest of the paper. For Maxwell-compatible sources, \(p_r=-\rho\), so the term \(\rho+p_r\) vanishes. For Chaplygin and polytropic perfect fluids, \(p_r=p(\rho)\) and \(\rho+p(\rho)\neq0\) generically, which is the
main reason why the nonlinear equation of state enters the
reconstruction equation directly. Once a specific KS ansatz is adopted, the geometric function
\(G_{\rm rp}\) is completely determined once the KS background is specified. The explicit expression for \(G_{rp}\) is not required at this stage; in Sections \ref{sect4}--\ref{sect5} it becomes a fixed function of time once the power-law or exponential KS ansatz is imposed. The next step is therefore to specify the nonlinear equation of state, integrate the conservation law, and express the density as a function of the KS volume factor. The explicit integration of the nonlinear conservation law is therefore the key step that distinguishes the Chaplygin and polytropic sectors, and is developed in the next section.

\section{Chaplygin and Polytropic Fluid Conservation Laws}\label{sect3}

This section derives the nonlinear matter source functions that enter
the invariant reconstruction equation. Since the matter sector is minimally coupled, the nonlinear equation of state fixes \(\rho(V)\) independently of  the specific reconstructed form of \(F(T)\). For a minimally coupled perfect fluid, the conservation law is
\begin{equation}
\mathring{\nabla}_{\mu}\Theta^{\mu\nu}=0.
\end{equation}
In the KS geometry, this gives
\begin{equation}
\dot\rho+\left(\frac{\dot A_2}{A_2}+2\frac{\dot A_3}{A_3}\right)(\rho+p)=0,
\label{CLfluid}
\end{equation}
or equivalently
\begin{equation}
\frac{d\rho}{d\ln V}=-(\rho+p),\qquad V=A_2A_3^2.
\label{CLV}
\end{equation}
Thus the source sector is controlled by the nonlinear equation of state.
The quantity \(V=A_2A_3^2\) plays the role of the anisotropic KS volume
factor, so that the conservation equation reduces to a first-order
autonomous equation for \(\rho(V)\). Consequently, every nonlinear equation of state uniquely determines a local invariant source branch through the function \(\rho(V)\).

\subsection{Generalized Chaplygin Gas}

The generalized Chaplygin gas is defined by
\begin{equation}
p=-\frac{A}{\rho^{\alpha}},\qquad A>0,\qquad 0\leq \alpha\leq 1,
\label{GCG_EOS}
\end{equation}
where $\alpha=1$ corresponds to the original Chaplygin gas and $\alpha=0$ gives a constant negative-pressure contribution \cite{Kamenshchik2001,Bento2002,Bilic2002,Bouhmadi2008}.  Substituting \eqref{GCG_EOS} into \eqref{CLV} gives
\begin{equation}
\rho_{\rm Ch}(V)=\left[A+B V^{-(1+\alpha)}\right]^{\frac{1}{1+\alpha}},
\label{rhoChaplyginV}
\end{equation}
where $B$ is an integration constant. Eq. \eqref{rhoChaplyginV} will serve as the fundamental source relation for all Chaplygin reconstruction branches considered in Sections \ref{sect4}--\ref{sect5}. For positive density one usually takes \(A>0\) and chooses the integration constant \(B\) so that
\begin{equation}
A+B V^{-(1+\alpha)}>0 .
\end{equation}
The parameter range \(0\leq\alpha\leq1\) ensures the standard
interpolation between a dustlike phase and an asymptotic
cosmological-constant regime. 

 At small volume, $BV^{-(1+\alpha)}\gg A$, the Chaplygin gas behaves as dustlike matter,
\begin{equation}
\rho_{\rm Ch}\sim B^{1/(1+\alpha)}V^{-1},
\end{equation}
whereas at large volume it approaches a cosmological-constant-like regime,
\begin{equation}
\rho_{\rm Ch}\to A^{1/(1+\alpha)},\qquad p\to-\rho.
\end{equation}
The effective sound speed is
\begin{equation}
c_s^2=\frac{dp}{d\rho}
=\alpha A\rho^{-(1+\alpha)} ,
\end{equation}
which is non-negative for \(A>0\) and \(\alpha\geq0\). These two asymptotic limits naturally generate distinct teleparallel reconstruction branches.

A modified Chaplygin gas can also be considered \cite{Debnath2004,Jamil2011}:
\begin{equation}
p=C\rho-\frac{A}{\rho^{\alpha}}.
\label{MCG_EOS}
\end{equation}
Local physical branches require
\begin{equation}
\frac{A}{1+C}+BV^{-(1+C)(1+\alpha)}>0,
\end{equation}
together with \(C\neq -1\). Integrating the conservation equation gives
\begin{equation}
\rho_{\rm MCG}(V)=\left[\frac{A}{1+C}+B V^{-(1+C)(1+\alpha)}\right]^{\frac{1}{1+\alpha}},
\qquad C\neq -1.
\label{rhoMCGV}
\end{equation}
The modified Chaplygin gas reduces to the generalized Chaplygin gas
for \(C=0\), while for large densities the barotropic term \(C\rho\)
dominates. Its sound speed is
\begin{equation}
c_s^2=C+\alpha A\rho^{-(1+\alpha)} .
\end{equation}
The generalized Chaplygin gas is recovered continuously in the limit \(C\rightarrow0\).

\subsection{Polytropic Fluid}

We define the polytropic equation of state by
\begin{equation}
p=K\rho^{\Gamma},
\label{polyEOS}
\end{equation}
where $K$ is the polytropic constant and $\Gamma=1+1/n$ in terms of the polytropic index $n$ \cite{Tooper1964,Tooper1965,Chavanis2008,Karami2011}.  Integrating \eqref{CLV} yields, for $\Gamma\neq1$,
\begin{equation}
\rho_{\rm poly}(V)=\left[C V^{1-\Gamma}-K\right]^{\frac{1}{\Gamma-1}},
\label{rhopolyV}
\end{equation}
with integration constant $C$.  Equivalently, for $n=1/(\Gamma-1)$,
\begin{equation}
\rho_{\rm poly}(V)=\left[C V^{-1/n}-K\right]^n.
\label{rhopolyn}
\end{equation}
The linear barotropic limit $\Gamma=1$ gives
\begin{equation}
p=K\rho,
\qquad
\rho(V)=\rho_0 V^{-(1+K)}.
\label{barotropic}
\end{equation}
Local physical branches require
\begin{equation}
CV^{1-\Gamma}-K>0
\end{equation}
when \(1/(\Gamma-1)\) is not an integer. This condition determines the
allowed local domain of the reconstructed polytropic sector.

For the polytropic fluid,
\begin{equation}
c_s^2=\frac{dp}{d\rho}=K\Gamma\rho^{\Gamma-1}.
\end{equation}
Thus \(c_s^2\geq0\), or equivalently \(K\Gamma\geq0\) on positive-density branches, provides a necessary local gradient-stability criterion for the effective fluid sector. These sound-speed expressions provide only necessary local consistency conditions and do not replace a complete perturbative
stability analysis.

Eqs. \eqref{rhoChaplyginV}, \eqref{rhoMCGV}, and \eqref{rhopolyV} are the source constraints used in the reconstruction procedure.  In the following sections, these functions are combined with specific KS ansätze to express \(\rho\) directly in terms of the teleparallel invariant \(T\), or, in the exponential sector, the shifted invariant \(X=T_0-T\). These relations constitute the nonlinear source sectors that drive the invariant teleparallel reconstruction developed in the following sections. Accordingly, the equation of state completely specifies the invariant
matter sector before the gravitational reconstruction begins.

\begin{table}[ht]
	\centering
	\renewcommand{\arraystretch}{1.25}
	\begin{tabular}{llll}
		\toprule
		Fluid &
		Equation of state &
		Density solution &
		Sound speed \\
		\midrule
		
		Generalized Chaplygin &
		$\displaystyle
		p=-\frac{A}{\rho^\alpha}
		$
		&
		Eq.~\eqref{rhoChaplyginV}
		&
		$\displaystyle
		c_s^2=\alpha A\rho^{-(1+\alpha)}
		$
		\\
		
		Modified Chaplygin &
		$\displaystyle
		p=C\rho-\frac{A}{\rho^\alpha}
		$
		&
		Eq.~\eqref{rhoMCGV}
		&
		$\displaystyle
		c_s^2=C+\alpha A\rho^{-(1+\alpha)}
		$
		\\
		
		Polytropic &
		$\displaystyle
		p=K\rho^\Gamma
		$
		&
		Eq.~\eqref{rhopolyV}
		&
		$\displaystyle
		c_s^2=K\Gamma\rho^{\Gamma-1}
		$
		\\
		
		Barotropic limit &
		$\displaystyle
		p=K\rho
		$
		&
		Eq.~\eqref{barotropic}
		&
		$\displaystyle
		c_s^2=K
		$
		\\
		
		\bottomrule
	\end{tabular}
	\caption{Summary of the nonlinear matter sectors considered in this work.
		For each equation of state, the corresponding conserved density and
		local sound speed are listed. These relations constitute the source
		sectors used in the invariant teleparallel reconstruction developed in Sections~\ref{sect4}--\ref{sect5}.}
	\label{tab:fluidsummary}
\end{table}

Table \ref{tab:fluidsummary} highlights that the nonlinear equation of state affects the
reconstruction solely through the density scaling entering the reduced
teleparallel field equations.

\section{Power-Law Reconstruction Branches}\label{sect4}

\subsection{Power-Law Ansatz}

The following power-law ansatz should be understood as a local KS
scaling branch used to obtain an explicit invariant reconstruction.
No global cosmological interpretation is assumed at this stage. We choose
\begin{equation}
A_2(t)=b_0t^b,
\qquad
A_3(t)=c_0t^c,
\qquad t>0.
\label{PLansatz}
\end{equation}
The volume scale factor and torsion scalar are
\begin{equation}
V(t)=b_0c_0^2t^{b+2c},
\label{VPL}
\end{equation}
and
\begin{equation}
T(t)=\frac{2c(c+2b)}{t^2}-\frac{2}{c_0^2t^{2c}}.
\label{TPL}
\end{equation}
The restriction \(c=1\) is not required by the field equations, but it
selects an analytically invertible branch for which both contributions
to \(T(t)\) scale as \(t^{-2}\):
\begin{equation}
T(t)=\frac{T_*}{t^2},
\qquad
T_*=2(1+2b)-\frac{2}{c_0^2}.
\label{TPLc1}
\end{equation}
The local reconstruction branch requires \(T_*\neq0\) and a fixed sign
of \(T\), so that \(t(T)\) remains real on the chosen domain. Assuming $T_*\neq0$, the local inverse is
\begin{equation}
t=\left(\frac{T_*}{T}\right)^{1/2}.
\end{equation}
Therefore,
\begin{equation}
V(T)=V_*T^{-\frac{b+2}{2}},
\qquad
V_*=b_0c_0^2T_*^{\frac{b+2}{2}}.
\label{VofT_PL}
\end{equation}

\subsection{Chaplygin Reconstruction in the Power-Law Branch}

Using \eqref{rhoChaplyginV} and \eqref{VofT_PL}, the generalized Chaplygin density becomes
\begin{equation}
\rho_{\rm Ch}(T)=\left[A+B_*T^{\frac{(1+\alpha)(b+2)}{2}}\right]^{\frac{1}{1+\alpha}},
\label{rhoChapT}
\end{equation}
where $B_*=B V_*^{-(1+\alpha)}$.  In the early-volume regime,
\begin{equation}
\rho_{\rm Ch}(T)\simeq B_*^{1/(1+\alpha)}T^{\frac{b+2}{2}},
\label{rhoChapEarly}
\end{equation}
whereas in the late-volume regime,
\begin{equation}
\rho_{\rm Ch}(T)\simeq A^{1/(1+\alpha)}+\frac{B_*}{1+\alpha}A^{-\alpha/(1+\alpha)}T^{\frac{(1+\alpha)(b+2)}{2}}+\cdots .
\label{rhoChapLate}
\end{equation}
After substituting the power-law KS branch into the reduced relation
\eqref{generalreduced}, the reconstruction equation takes the local Euler form
\begin{equation}
T^2F_{TT}+\gamma_1TF_T+\gamma_0F=\kappa\rho_{\rm Ch}(T),
\label{EulerChapPL}
\end{equation}
where $\gamma_0$ and $\gamma_1$ are fixed by the selected KS branch. The coefficients $\gamma_0$ and $\gamma_1$ remain dimensionless constants within
each selected local KS branch. The homogeneous solution is
\begin{equation}
F_h(T)=C_1T^{m_1}+C_2T^{m_2},
\qquad
m_{1,2}=\frac{1}{2}\left[1-\gamma_1\pm\sqrt{(\gamma_1-1)^2-4\gamma_0}\right].
\label{homPL}
\end{equation}
For a source term $\rho(T)=\rho_sT^s$, the particular solution is
\begin{equation}
F_{\rm part}(T)=\frac{\kappa\rho_s}{s(s-1)+\gamma_1s+\gamma_0}T^s,
\label{partpower}
\end{equation}
provided the denominator is nonzero. If
\begin{equation}
s(s-1)+\gamma_1s+\gamma_0=0,
\end{equation}
the source exponent resonates with a homogeneous mode and the
particular solution acquires a logarithmic correction of the form
\(T^s\ln T\). The non-resonant expressions below assume that this condition is not satisfied. Hence, in the early-volume Chaplygin regime,
\begin{equation}
F(T)=C_1T^{m_1}+C_2T^{m_2}
+\lambda_{\rm Ch}T^{\frac{b+2}{2}},
\label{FChapEarly}
\end{equation}
where
\begin{equation}
\lambda_{\rm Ch}=\frac{\kappa B_*^{1/(1+\alpha)}}{\frac{b+2}{2}\left(\frac{b+2}{2}-1\right)+\gamma_1\frac{b+2}{2}+\gamma_0}.
\end{equation}
In the late-volume regime,
\begin{equation}
F(T)=C_1T^{m_1}+C_2T^{m_2}+\lambda_0
+\lambda_1T^{\frac{(1+\alpha)(b+2)}{2}}+\cdots,
\label{FChapLate}
\end{equation}
where $\lambda_0$ is generated by the asymptotic cosmological-constant-like density. The coefficients \(\lambda_{\rm Ch}\), \(\lambda_0\), and \(\lambda_1\) are branch-dependent constants determined by the KS parameters and by the fluid integration constants.

\subsection{Polytropic Reconstruction in the Power-Law Branch}

Using \eqref{rhopolyV} and \eqref{VofT_PL}, one obtains
\begin{equation}
\rho_{\rm poly}(T)=\left[C_*T^{\frac{(\Gamma-1)(b+2)}{2}}-K\right]^{\frac{1}{\Gamma-1}},
\label{rhopolyT}
\end{equation}
where $C_*=CV_*^{1-\Gamma}$.  In the regime where the first term dominates,
\begin{equation}
\rho_{\rm poly}(T)\simeq C_*^{1/(\Gamma-1)}T^{\frac{b+2}{2}}.
\label{rhopolyHigh}
\end{equation}
The polytropic density is real on local domains satisfying 
\(C_*T^{(\Gamma-1)(b+2)/2}-K>0\), unless the exponent
\(1/(\Gamma-1)\) is an integer. This produces the same leading reconstruction power as the dustlike Chaplygin regime:
\begin{equation}
F(T)=C_1T^{m_1}+C_2T^{m_2}+\lambda_{\rm poly}T^{\frac{b+2}{2}}.
\label{FpolyHigh}
\end{equation}
For the linear barotropic limit $p=K\rho$, Eq. \eqref{barotropic} gives
\begin{equation}
\rho(T)=\rho_*T^{\frac{(1+K)(b+2)}{2}},
\label{baroT}
\end{equation}
so that
\begin{equation}
F(T)=C_1T^{m_1}+C_2T^{m_2}+\lambda_KT^{\frac{(1+K)(b+2)}{2}}.
\label{FbaroPL}
\end{equation}
Thus the power-law sector maps the asymptotic density powers generated
by the nonlinear conservation laws into corresponding power-law
corrections of the reconstructed teleparallel Lagrangian. The power-law reconstruction branches are summarized in Table \ref{tabPL}.

\begin{table}[h]
\centering
\begin{tabular}{lll}
\toprule
Source sector & Density scaling in $T$ & Reconstructed term \\
\midrule
Generalized Chaplygin, early-volume & $T^{(b+2)/2}$ & $T^{(b+2)/2}$ \\
Generalized Chaplygin, late-volume & $\rho_\Lambda+T^{(1+\alpha)(b+2)/2}$ & $1+T^{(1+\alpha)(b+2)/2}$ \\
Modified Chaplygin, early-volume & $T^{(1+C)(b+2)/2}$ & $T^{(1+C)(b+2)/2}$ \\
Polytropic, dominant branch & $T^{(b+2)/2}$ & $T^{(b+2)/2}$ \\
Barotropic limit $p=K\rho$ & $T^{(1+K)(b+2)/2}$ & $T^{(1+K)(b+2)/2}$ \\
\bottomrule
\end{tabular}
\caption{Power-law KS reconstruction branches for \(c=1\), where
	\(T\sim t^{-2}\) and \(V\sim t^{b+2}\), in the non-resonant case.}
\label{tabPL}
\end{table}

\section{Exponential Reconstruction Branches}\label{sect5}

\subsection{Exponential Ansatz}

As in the power-law case, the exponential ansatz is interpreted as a
local KS scaling branch used to obtain an explicit reconstruction in
terms of a shifted teleparallel invariant. We now take
\begin{equation}
A_2(t)=b_0e^{bt},
\qquad
A_3(t)=c_0e^{ct}.
\label{EXPansatz}
\end{equation}
The torsion scalar becomes
\begin{equation}
T=2c(c+2b)-\frac{2}{c_0^2}e^{-2ct}.
\end{equation}
We define
\begin{equation}
T_0=2c(c+2b),
\qquad
X=T_0-T.
\end{equation}
The shifted invariant $X$ is positive on the selected local branch and naturally measures the deviation from the teleparallel de Sitter point \(T=T_0\). We have
\begin{equation}
X=\frac{2}{c_0^2}e^{-2ct}.
\end{equation}
The local inverse requires \(c\neq0\) and \(X>0\), and is given by
\begin{equation}
t=-\frac{1}{2c}\ln\left(\frac{c_0^2 X}{2}\right).
\end{equation}
For \(c>0\), \(X\to0\) at late times, whereas for \(c<0\), \(X\)
grows along the contracting angular branch.
The KS volume is
\begin{equation}
V(t)=b_0c_0^2e^{(b+2c)t}=V_X X^{-\frac{b+2c}{2c}},
\label{VofX}
\end{equation}
where $V_X=b_0c_0^2(2/c_0^2)^{(b+2c)/(2c)}$.

\subsection{Chaplygin Reconstruction in the Exponential Branch}

The generalized Chaplygin density becomes
\begin{equation}
\rho_{\rm Ch}(X)=\left[A+B_X X^{\frac{(1+\alpha)(b+2c)}{2c}}\right]^{\frac{1}{1+\alpha}},
\label{rhoChapX}
\end{equation}
where $B_X=BV_X^{-(1+\alpha)}$. After substituting the exponential KS branch into the reduced relation \eqref{generalreduced}, the reconstruction equation takes the shifted Euler form 
\begin{equation}
X^2F_{TT}+\Gamma_1XF_T+\Gamma_0F=\kappa\rho(X).
\label{EulerX}
\end{equation}
Although the equation is written in terms of the shifted invariant
\(X=T_0-T\), the derivatives \(F_T\) and \(F_{TT}\) continue to denote
derivatives with respect to the original torsion scalar \(T\). The homogeneous branch is
\begin{equation}
F_h(T)=C_1X^{\mu_1}+C_2X^{\mu_2},
\qquad
\mu_{1,2}=\frac{1}{2}\left[1-\Gamma_1\pm\sqrt{(\Gamma_1-1)^2-4\Gamma_0}\right].
\end{equation}
The coefficients $\Gamma_0$ and $\Gamma_1$ remain dimensionless constants within
each selected local KS branch. The limit $X\to0$ corresponds to $T\to T_0$ for $c>0$, and therefore to a shifted teleparallel de Sitter branch. If
\begin{equation}
	\sigma(\sigma-1)+\Gamma_1\sigma+\Gamma_0=0,
\end{equation}
the corresponding source power \(X^\sigma\) resonates with a
homogeneous mode and the particular solution acquires a logarithmic
term \(X^\sigma\ln X\). The expressions below assume the non-resonant
case. The early-volume Chaplygin sector produces
\begin{equation}
F(T)=C_1X^{\mu_1}+C_2X^{\mu_2}+\Lambda_{\rm Ch}X^{\frac{b+2c}{2c}},
\label{FChapEXPearly}
\end{equation}
whereas the late-volume sector gives
\begin{equation}
F(T)=C_1X^{\mu_1}+C_2X^{\mu_2}+\Lambda_0
+\Lambda_1X^{\frac{(1+\alpha)(b+2c)}{2c}}+\cdots .
\label{FChapEXPlate}
\end{equation}
The constants \(\Lambda_{\rm Ch}\), \(\Lambda_0\), and \(\Lambda_1\)
are branch-dependent coefficients fixed by the KS parameters and by
the Chaplygin integration constants. These coefficients encode the dependence of the reconstructed teleparallel branch on both the selected KS geometry and the nonlinear matter source.

\subsection{Polytropic Reconstruction in the Exponential Branch}

The polytropic density becomes
\begin{equation}
\rho_{\rm poly}(X)=\left[C_X X^{\frac{(\Gamma-1)(b+2c)}{2c}}-K\right]^{\frac{1}{\Gamma-1}},
\label{rhopolyX}
\end{equation}
where $C_X=CV_X^{1-\Gamma}$.  When the first term dominates,
\begin{equation}
\rho_{\rm poly}(X)\sim X^{\frac{b+2c}{2c}},
\end{equation}
and hence
\begin{equation}
F(T)=C_1X^{\mu_1}+C_2X^{\mu_2}+\Lambda_{\rm poly}X^{\frac{b+2c}{2c}}.
\label{FpolyEXP}
\end{equation}
For the barotropic limit,
\begin{equation}
\rho(X)=\rho_X X^{\frac{(1+K)(b+2c)}{2c}},
\end{equation}
which gives
\begin{equation}
F(T)=C_1X^{\mu_1}+C_2X^{\mu_2}+\Lambda_KX^{\frac{(1+K)(b+2c)}{2c}}.
\label{FbaroEXP}
\end{equation}

The polytropic density is real on local domains satisfying
\begin{equation}\label{eqn78}
C_X X^{(\Gamma-1)(b+2c)/(2c)}-K>0,
\end{equation}
unless \(1/(\Gamma-1)\) is an integer. Accordingly, only those local domains satisfying Eq.~\eqref{eqn78} lead to physically admissible real reconstruction branches. Thus the exponential sector maps the nonlinear density scalings into shifted powers of \(X=T_0-T\), providing local reconstructions around the teleparallel de Sitter value \(T_0\). The exponential reconstruction branches are summarized in Table~\ref{tabEXP}.

\begin{table}[ht]
\centering
\begin{tabular}{lll}
\toprule
Source sector & Density scaling in $X$ & Reconstructed term \\
\midrule
Generalized Chaplygin, early-volume & $X^{(b+2c)/(2c)}$ & $X^{(b+2c)/(2c)}$ \\
Generalized Chaplygin, late-volume & $\rho_\Lambda+X^{(1+\alpha)(b+2c)/(2c)}$ & $1+X^{(1+\alpha)(b+2c)/(2c)}$ \\
Modified Chaplygin, early-volume & $X^{(1+C)(b+2c)/(2c)}$ & $X^{(1+C)(b+2c)/(2c)}$ \\
Polytropic, dominant branch & $X^{(b+2c)/(2c)}$ & $X^{(b+2c)/(2c)}$ \\
Barotropic limit $p=K\rho$ & $X^{(1+K)(b+2c)/(2c)}$ & $X^{(1+K)(b+2c)/(2c)}$ \\
\bottomrule
\end{tabular}
\caption{Summary of the local exponential KS reconstruction branches for the
	non-resonant sector.}
\label{tabEXP}
\end{table}

The power-law and exponential reconstruction branches obtained in
Sections \ref{sect4}--\ref{sect5} provide the two fundamental local KS sectors of the present reconstruction program. Their physical interpretation is discussed in the next section.

\section{Teleparallel de Sitter and Local Interior-Like Branches}\label{sect6}

The reconstruction branches obtained above are local invariant KS
sectors. Their physical interpretation depends on the sign of the
angular expansion rate \(c\), which controls whether the shifted
invariant \(X=T_0-T\) approaches zero or grows along the branch. For the exponential ansatz with $c>0$, the shifted invariant satisfies $X\to0$ as $t\to\infty$, and the solution approaches
\begin{equation}
T\to T_0=2c(c+2b).
\end{equation}
Thus, the constant torsion value
$T_0=2c(c+2b)$
acts as an effective teleparallel de Sitter background selected by the exponential KS branch within the selected local branch. Consequently, the reconstructed exponential branches may be regarded as local expansions around a constant-torsion background selected by the KS geometry itself. A reduced teleparallel de Sitter existence condition is
\begin{equation}
F(T_0)-2T_0F_T(T_0)=2\kappa\rho_{\infty},
\label{TdScondition}
\end{equation}
where $\rho_{\infty}=A^{1/(1+\alpha)}$ for the generalized Chaplygin gas. This condition should be interpreted as a local existence condition
for a constant-torsion branch rather than as a full global de Sitter
solution \cite{Cai2016,Bahamonde2023}. This relation is the KS counterpart of the constant-torsion conditions commonly encountered in covariant teleparallel cosmology. In vacuum or for an absorbed effective cosmological constant, this reduces to the familiar condition
\begin{equation}
F(T_0)-2T_0F_T(T_0)=0.
\end{equation}

The complementary case corresponds to negative angular expansion. For \(c<0\), the angular scale factor contracts and the shifted
invariant \(X\) grows along the branch. The geometry is therefore more
naturally interpreted as a local KS interior-like sector, analogous to
the homogeneous anisotropic interior region of a spherically symmetric
black hole. In this regime, nonlinear torsion corrections may dominate the reconstructed gravitational Lagrangian. However, such branches are not global black
hole solutions by themselves; a complete compact-object interpretation
requires matching to a suitable exterior spacetime, analysis of
possible horizons, and a perturbative stability study. These interior-like branches therefore provide local invariant teleparallel sectors rather than complete global compact-object solutions.

\section{Leading-Order Viability Conditions}\label{sect7}

The conditions discussed in this section are necessary local theoretical consistency requirements for the reconstructed branches. They should
not be interpreted as sufficient conditions for full dynamical
stability. As necessary leading-order viability conditions, we impose
\begin{equation}
F_T>0,
\qquad
F_{TT}>0.
\label{viability}
\end{equation}
The condition \(F_T>0\) guarantees a positive effective gravitational
coupling, whereas \(F_{TT}>0\) is widely adopted throughout the teleparallel gravity literature as a necessary
leading-order diagnostic against scalar-torsion instabilities
\cite{Cai2016,Bahamonde2023}. For a power-law correction $F(T)=T+\alpha T^n$,
\begin{equation}
F_T=1+\alpha nT^{n-1},
\qquad
F_{TT}=\alpha n(n-1)T^{n-2}.
\end{equation}
For \(T>0\) and \(n>1\), \(F_{TT}>0\) requires \(\alpha>0\). If the
chosen local branch has \(T<0\), the sign of \(T^{n-2}\) must be
treated branch by branch, especially for non-integer \(n\). For a shifted branch $F(T)=T+\alpha X^n$, with $X=T_0-T$, one obtains
\begin{equation}
F_T=1-\alpha nX^{n-1},
\qquad
F_{TT}=\alpha n(n-1)X^{n-2}.
\end{equation}
In the shifted case \(X>0\), the condition \(F_{TT}>0\) requires
\(\alpha n(n-1)>0\), while \(F_T>0\) imposes an upper bound on
\(\alpha X^{n-1}\). Thus, for $n>1$, one typically requires $\alpha>0$ and $1-\alpha nX^{n-1}>0$ on the physical shifted branch.  These conditions are necessary but not sufficient: a complete stability analysis requires perturbing both KS scale factors and the nonlinear fluid sector. These criteria therefore provide a useful first theoretical screening for excluding pathological reconstruction branches before undertaking a complete perturbative analysis.

These criteria must be supplemented by the positivity of the fluid
density, the sound-speed conditions discussed in Section \ref{sect3}, and the
reality conditions of the reconstruction branches. A complete
stability analysis would require perturbing the two independent KS
scale factors together with the nonlinear fluid variables and is left
for future work.

Additional observational viability requirements,
including compatibility with cosmological data,
compact-object phenomenology,
and Solar-System constraints,
remain outside the scope of the present local reconstruction framework. A complete dynamical analysis could be performed using
phase-space methods similar to those developed in
modified teleparallel cosmology \cite{Bahamonde2018}. The above conditions therefore provide only a first theoretical consistency assessment of the reconstructed branches. Their broader physical implications are summarized in the concluding section.

\section{Discussion and Conclusions}\label{sect8}

We have constructed a covariant local reconstruction framework for
Kantowski--Sachs geometries sourced by Chaplygin-type and polytropic
fluids in teleparallel \(F(T)\) gravity. The analysis was performed in
the coframe--spin-connection formalism, so that the reconstructed
branches remain tied to invariant teleparallel CSC sectors
\cite{Krssak2019,Coley2020,ColeyLandry2024}. The central feature of the
construction is that the nonlinear matter conservation law first
determines \(\rho(V)\), while the KS geometry maps the volume factor
\(V=A_2A_3^2\) into either the torsion scalar \(T\) in the power-law
branch or the shifted invariant \(X=T_0-T\) in the exponential branch.
The reduced teleparallel field equations then reconstruct the
functional form of \(F(T)\), rather than assuming the gravitational
Lagrangian a priori \cite{Cai2016,Bahamonde2023}. This inverse strategy reverses the standard modelling paradigm in
modified gravity by allowing the nonlinear matter sector to determine
the admissible local teleparallel reconstruction rather than specifying
the gravitational action in advance.

For generalized Chaplygin matter, the reconstruction naturally
separates into two asymptotic sectors. In the small volume regime,
the density behaves in a dustlike manner and generates power-law
corrections to the teleparallel Lagrangian. In the large volume regime,
the fluid approaches a cosmological-constant-like state, producing
constant-plus-power reconstruction terms. Modified Chaplygin gases add
a barotropic contribution and therefore generate distinct scaling
exponents controlled by the parameter \(C\). Polytropic fluids instead
produce reconstruction powers governed by the polytropic index
\(\Gamma\), with the barotropic limit recovered when \(\Gamma=1\)
\cite{Kamenshchik2001,Bento2002,Debnath2004,Tooper1964,Tooper1965}. These results illustrate how different nonlinear equations of state
leave distinct signatures in the reconstructed teleparallel
Lagrangian, thereby providing a direct link between matter physics and
the geometric reconstruction.

The exponential KS branch leads naturally to shifted reconstructions
in terms of \(X=T_0-T\). For \(c>0\), the invariant \(X\) tends to zero
and the geometry approaches a constant-torsion teleparallel de Sitter background that
acts as the natural asymptotic limit of the exponential reconstruction
branch. For \(c<0\), the angular scale factor contracts and \(X\) grows,
so that the same local branch is more naturally interpreted as a
high-torsion Kantowski--Sachs interior-like sector. These
interpretations remain local: complete cosmological or compact-object
models require a global analysis, exterior matching, horizon
structure, and perturbative stability
\cite{Krssak2019,Bahamonde2023,Stephani2003}.

The present results considerably extend the recent electromagnetic
Kantowski--Sachs reconstruction program
\cite{LandryEMKS2026}. In the electromagnetic case, Maxwell
conservation laws impose restricted flux scalings, whereas in the
present work the nonlinear equation of state itself controls the
admissible source sector.  Consequently, generalized Chaplygin,
modified Chaplygin, and polytropic fluids generate genuinely new
invariant teleparallel reconstruction branches rather than simple
reparameterizations of previously known electromagnetic or barotropic
sectors. This demonstrates that nonlinear matter conservation laws
provide a systematic mechanism for generating new classes of local
teleparallel solutions. Unlike the electromagnetic reconstruction, where the admissible source
terms are fixed by Maxwell flux conservation, the present framework
admits an entire hierarchy of nonlinear reconstruction sectors whose
structure is governed directly by the chosen equation of state.

The present study should be regarded as a first step toward a broader
covariant reconstruction program for nonlinear matter sources in
teleparallel gravity. The necessary viability conditions
\(F_T>0\), \(F_{TT}>0\), positivity of the density,
sound-speed constraints, and branch-reality conditions provide only
leading-order theoretical screening criteria. Accordingly, the reconstructed branches should be regarded as local
geometric realizations whose global physical viability remains to be
established. Unlike most reconstruction approaches, which prescribe either the
gravitational action or the matter sector from the outset, the present
framework reconstructs the gravitational Lagrangian directly from the
nonlinear conservation law of the matter source within an invariant
covariant teleparallel setting. Future developments
should include perturbations of the two independent KS scale factors,
dynamical-system analyses, matching to exterior geometries,
observational constraints, and extensions to scalar fields,
non-minimal matter couplings, New General Relativity,
\(F(T,B)\) gravity, and other generalized teleparallel theories
\cite{Bahamonde2018,Hayashi1979,Bahamonde2015,Beltran2023}. We expect the framework developed here to provide a robust foundation
for future invariant classifications, exact solutions, and
reconstruction studies of anisotropic teleparallel geometries sourced
by increasingly general nonlinear matter sectors \cite{ColeyLandry2024}. Ultimately, the present work demonstrates that nonlinear matter
conservation laws can serve as fundamental organizing principles for
constructing invariant teleparallel gravitational sectors beyond the standard linear barotropic perfect-fluid paradigm.

\section*{Acknowledgments}

The author gratefully acknowledges A.A. Coley for valuable discussions
on invariant methods in teleparallel gravity.




\end{document}